\documentclass[12pt]{article}
\usepackage{amssymb,amsmath}
\usepackage[noblocks]{authblk}
\usepackage[top=0.75in, bottom=0.75in, left=0.75in, right=0.75in, dvips]{geometry}
\usepackage{caption}
\begin{document}
\textwidth 10.0in 
\textheight 9.0in 
\topmargin -0.60in
\title{Covariant Gauge Fixing and Canonical Quantization}
\author[1,2]{D.G.C. McKeon}
\affil[1] {Department of Applied Mathematics, The
University of Western Ontario, London, ON N6A 5B7, Canada} 
\affil[2] {Department of Mathematics and
Computer Science, Algoma University, Sault St.Marie, ON P6A
2G4, Canada}

\maketitle

\maketitle

\begin{abstract}
Theories that contain first class constraints possess gauge invariance which results in the necessity of altering the measure in the associated quantum mechanical path integral.  If the path integral is derived from the canonical structure of the theory, then the choice of gauge conditions used in constructing Faddeev's measure cannot be covariant. This shortcoming is normally overcome either by using the ``Faddeev-Popov'' quantization procedure, or by the approach of Batalin-Fradkin-Fradkina-Vilkovisky, 
and then demonstrating that these approaches are equivalent to the path integral constructed from the canonical approach with Faddeev's measure. We propose in this paper an alternate way of defining the measure for the path integral when it is constructed using the canonical procedure for theories containing first class constraints and that this new approach can be used in conjunction with covariant gauges. This procedure follows the Faddeev-Popov approach, but rather than working with the form of the gauge transformation in configuration space, it employs the generator of the gauge transformation in phase space. We demonstrate this approach to the path integral by applying it to Yang-Mills theory, a spin-two field and the first order Einstein-Hilbert action in two dimensions. The problems associated with defining the measure for theories containing second-class constraints and ones in which there are fewer secondary first class constraints than primary first class constraints are discussed.
\end{abstract}

\section{Introduction}

The canonical quantization procedure involves selection of a particular coordinate system. This leads to a breaking of the manifest covariance present in a model. Stueckelberg [1] was the first to overcome this shortcoming in quantum electrodynamics; subsequently Feynman [2], Schwinger [3] and Tomonaga [4] addressed this issue.

When the techniques which had worked so well in quantum electrodynamics were applied to non-Abelian gauge theories, Feynman found that at one loop order a Fermionic scalar ``ghost'' field had to contribute for the results to be consistent with unitarity [5]; a general discussion of these ghost fields was given by DeWitt [6], Mandelstam [7] and Faddeev and Popov [8]. This last approach (FP) involved modifying the standard path integral for the generating function $Z[J]$
\[
Z[J] = \int D\Phi \exp i \int dx (\mathcal{L} (\Phi) + J \Phi)\eqno(1a)
\]
(where $\mathcal{L}$ is the Lagrangian for the field $\Phi$) when a gauge invariance is present so that it becomes
\[
Z[J] = \int D\Phi \Delta_{FP} (\Phi) \exp i \int dx (\mathcal{L} (\Phi) - \frac{1}{2\alpha} (\chi (\Phi))^2 + J\Phi).\eqno(1b)
\]
In eq. (2), $\chi(\Phi)$ is the gauge condition that has been chosen and $\Delta_{FP}(\Phi)$ is the associated ``Faddeev-Popov determinant''. $Z[J]$ is independent of the choice of $\chi$. In appendix A, the means by which we can pass from eq. (1a) to eq. (1b) when considering a gauge theory is outlined in detail. 

This approach starts from the assumption that the path integral of eq. (1a) can be derived from the canonical quantization procedure and that if constraints were to arise in the canonical analysis of a model, then the associated gauge invariance that is present leads to divergent integration over physically equivalent configurations of the field $\Phi$. This divergence is ``factored out'' of the path integral in a way that is consistent with unitarity by systematically insertion of a gauge fixing condition $\chi (\Phi) = 0$ and accompanying this with the determinant $\Delta_{FP}$. However, the path integral of eq. (1a) does not follow automatically from canonical quantization; in fact canonical quantization leads to the path integral [9]
\[
Z[J] = \int D\Phi \,D\Pi \exp i \int dx \left( \Phi \frac{\partial\Pi}{\partial t} - \mathcal{H}_c (\Phi , \Pi) + J\Phi\right)\eqno(2)
\]
where $\Pi$ is the canonical momentum conjugate to $\Phi$ and $\mathcal{H}_c$ is the canonical Hamiltonian for the system. 

If $\mathcal{H}_c$ is quadratic in $\Pi$, then upon integration over $\Pi$, eq. (2) reduces to eq. (1a). Eq. (2) follows from the usual equation for the evolution of a dynamical variable $A(\Phi, \Pi)$
\[
\frac{dA}{dt} = \left\lbrace A, H_c \right\rbrace \nonumber \]
where $\left\lbrace , \right\rbrace$ denotes the Poisson Bracket (PB) and $H_c = \int dx\; \mathcal{H}_c$.  If the system has a set of primary constraints $\gamma_{a_{1}}$, then the Lagrangian equations of motion are equivalent to the equations
\[
\frac{dA}{dt} = \left\lbrace A, H_T \right\rbrace \quad , \quad \gamma_{a_{1}} = 0 \nonumber \]
in phase space, where $H_T = \int dx\; \mathcal{H}_T = \int dx (\mathcal{H}_c + \lambda_{a_{1}}\gamma_{a_{1}})$ is the ``total'' Hamiltonian.  This leads to the path integral 
\[
Z[J] = \int D \Phi \; D\Pi \; D\lambda_{a_{1}} \exp i \int dx \left( \Phi \frac{\partial\Pi}{\partial t} - \mathcal{H}_T (\Phi, \Pi) + J\Phi \right).\eqno(3)\]
However, the presence of first class constraints means that the total action in phase space\linebreak $S_T = \int dx \left( \Phi \frac{\partial \Pi}{\partial t} - \mathcal{H}_T (\Phi, \Pi)\right)$ possesses a gauge invariance, and hence the path integral in eq. (3) is ill defined. This is much like having the configuration space path integral of eq. (1a) being ill defined if $\int dx \mathcal{L}$ possesses a gauge invariance.

If the model being considered has $N$ generations of first class constraints $\phi_a(\Phi,\Pi)$, then Faddeev [10], by elimination of non-dynamical degrees of freedom has shown that the path integral 
\[
\hspace{-3cm} Z[J] = \int D\Phi\, D\Pi\, D\lambda_a \,\det \left\lbrace \phi_a, \chi_b\right\rbrace \delta(\chi_b)\eqno(4)\]
\[ \hspace{1.2cm} \times \exp i \int dx \left(\Phi \frac{\partial\Pi}{\partial t} - \mathcal{H}_c (\Phi, \Pi) - \lambda_a\phi_a(\Phi ,\Pi) + J\Phi\right)\nonumber
\]
(where $\chi_b$ is the gauge condition associated with $\phi_b$ and $\lambda_a$ are Lagrange multipliers) is identical to the path integral of eq. (2) provided in eq. (2) $\Phi$ and $\Pi$ are restricted to $\Phi^*$ and $\Pi^*$ which are the true dynamical degrees of freedom in the system. The presence of the determinant of the Poisson Bracket (PB) of $\phi_a$ and $\chi_b$ obscures how one could choose a gauge condition $\chi_b$ that is covariant, or how $Z[J]$ itself could be covariant, when the initial Lagrangian is manifestly covariant. Faddeev showed [10,11] that for Yang-Mills theory, eq. (4) is equivalent to eq. (1b) and so $Z[J]$ can be expressed in manifestly covariant form with a covariant gauge choice $\chi$.  However, it is not apparent that this demonstration can be extended to show that eqs. (1b) and (4) are equivalent for all models containing first class constraints.

In a series of papers, Batalin, Fradkin, Fradkina and Vilkovisky [12-18] (for reviews see refs. [19-27]) have shown how the BRST symmetry [28,29] associated with the path integral of eq. (1b) can be exploited to deal with the path integral both in configuration space and phase space in a way that can accommodate the most general structure of gauge algebra and of constraints.  This approach uses anti-commuting Grassmann fields to cancel the contribution to the path integral of non-physical (``gauge'') degrees of freedom that are present in the initial Lagrangian.  Covariant gauges can be incorporated into their discussion of the path integral of eq. (4) when first class constraints are present upon treating the Lagrange multipliers as dynamical variables. For Yang-Mills theory, this involves having to treat the temporal component of the vector field as being a Lagrange multiplier rather than a dynamical field.

Senjanovic [30] (see also Fradkin [31]) extended the approach used by Faddeev to derive eq. (4) to incorporate models containing second class constraints $\theta_a(\Phi,\Pi)$. The resulting path integral is 
\[
Z[J] = \int D\Phi\, D\Pi\, D\lambda_a\, D\kappa_a \det \left\lbrace \phi_a, \chi_b\right\rbrace \mathrm{\det}^{1/2}\left\lbrace \theta_a, \theta_b\right\rbrace  \delta(\chi_b)\nonumber \]
\[\times \exp i \int dx \bigg(\Pi \frac{\partial}{\partial t}\Phi - \mathcal{H}_c (\Phi, \Pi) - \lambda_a\phi_a(\Phi ,\Pi) \nonumber \]
\[ \hspace{1cm}- \kappa_a \theta_a (\Phi, \Pi) +  J\Phi\bigg), \eqno(5)
\]
where $\kappa_a$ is another set of Lagrange multiplier fields. In many cases, such as the Proca model [30] or Yang-Mills theories on a light-cone [32] the contribution of $\det^{1/2}\left\lbrace \theta_a, \theta_b\right\rbrace$ to the measure in eq. (5) is an innocuous constant. However, if it is not a constant (for example the model of refs. [33-37]) then it is not readily apparent how this contribution to the measure of eq. (5) can be made manifestly covariant [15].  There are a variety of techniques that in certain instances can be used to replace the original model in which second class constraints occur to an equivalent model in which only first class constraints appear [38-52].  This would permit one to quantize the model using the path integral of eq. (4). However, these procedures for eliminating second class constraints are often either impractical or even not feasible.

In this paper we propose an alternate way to quantize models containing first class constraints by using eq. (3). Rather than using the constraints and their associated gauge conditions to eliminate non-physical degrees of freedom, and then reverting to the original degrees of freedom in phase space (as was done by Faddeev [10] to obtain eq. (4)) we integrate over ``gauge orbits'' in phase space in eq. (3) as was done by Faddeev and Popov [8] in conjunction with the configuration space path integral of eq. (1a) to derive eq. (1b). The advantage of this approach is that one can immediately employ manifestly covariant gauge conditions in conjunction with the phase space path integral without promoting the Lagrange multipliers associated with the first class constraints to being dynamical variables, as was done in ref. [12]. One uses the generator of a gauge transformation in phase space in order to carry out this procedure [53-56].

We will first outline the general procedure we use to produce a path integral in phase space that employs an integral over ``gauge orbits'' that can be absorbed into an overall multiplicative factor. This technique is then applied to Yang-Mills theory, to the spin-two Lagrangian, to some aspects of the Einstein-Hilbert action in four dimensions and to the first order Einstein-Hilbert action in two dimensions. The implication of this approach to systems with second class constraints $\theta_a$ when $\left\lbrace \theta_a, \theta_b \right\rbrace$ is non-trivial is then discussed.  We also consider systems in which the number of primary first class constraints exceeds the number of secondary first class constraints. In appendix A we review the Faddeev-Popov procedures used in deriving eq. (1b). The generators of a gauge transformations in phase space are derived in appendix B. In appendix C the conversion of the path integral from phase space to configurations space is outlined.

\section{Covariant Gauge Fixing}

The presence of a gauge invariance in a model leaves the path integral of eq. (3) ill-defined. Rather than using the constraints present to eliminate non-physical degrees of freedom as in refs. [10,30], we will follow the approach of ref. [8] and insert a constant factor into this path integral in order that it become well-defined. In analogy with eq. (A.4a) we use the constant factor
\[
1 = \int D\mu_{a_{N}} \delta(\psi + \left\lbrace \psi , G\right\rbrace - k)\Delta . \eqno(6)
\]
In eq. (6), $\psi(\Phi ,\partial_\mu\Phi)$ is a (possibly covariant) gauge fixing functional, $G$ is the generator of eq. (B.2) used to generate gauge transformations with $\mu_{a_{N}}$ being associated with the last generation of primary first class constraints. The coefficients $\mu_{a_{1}} \ldots \mu_{a_{N-1}}$ appearing in $G$ that are associated with lower generation constraints are taken to be determined in terms of $\mu_{a_{N}}$ by eq. (B.5).  The factor of $\Delta$ in eq. (6) is a functional determinant, analogous to the Faddeev-Popov determinant $\det(\mathbf{FM})$ in eq. (A.4a), that ensures that the right side of eq. (6) is a constant. The field $k(x)$ is independent of the dynamical variables.

We can illustrate how this technique works by applying it to Yang-Mills theory for a vector field $A_\mu^a$.  With the metric $(+ + + -)$ and the Lagrangian
\[
\mathcal{L} = - \frac{1}{4} F_{\mu\nu}^a (A) F^{a\mu\nu}(A) \eqno(7)
\]
($D_\mu^{ab} = \partial_\mu\delta^{ab} + f^{apb}A_\mu^p$, $[D_\mu ,D_\nu]^{ab} = f^{apb}F_{\mu\nu}^p$) we see that the canonical momentum 
\[
\pi^{a\mu} = \frac{\partial\mathcal{L}}{\partial(\partial_0A_\mu^a)}\eqno(8)
\]
is given by 
\[
\pi^{a0} = 0\qquad \pi^{ai} = F_{0i}^a . \eqno(9,10)
\]
We thus have the primary constraint 
\[
 \phi_1^a = \pi^{a0} \eqno(11a)
\]
from which follows the secondary constraint
\[ \phi_2^a = D_i^{ab} \pi^{bi} .\eqno(11b)\]
Both of these constraints are first class and there are no third generation (tertiary) constraints. With the generator
\[ G = \int d^3x \left[ \mu_1^a \phi_1^a + \mu_2^a \phi_2^a\right] \eqno(12) \]
it follows from eq. (B.5) that 
\[ \mu_1^a = -D_0^{ab} \mu_2^b .\eqno(13)  \]
From eqs. (12,13,B.3) we see that the Lagrangian $\mathcal{L}$ of eq. (8) is invariant under the transformation\
\[ \delta A_\lambda^a = \left\lbrace A_{\lambda ,}^a (-D_0^{ab} \mu_2^b) \pi^{a0} + \mu_2^a (D_i^{ab} \pi^{bi})\right\rbrace \eqno(14) \]
\[\hspace{-3.7cm} = -D_\lambda^{ab} \mu_2^b \eqno(15) \]
which is the usual Yang-Mills gauge transformation.

We can now consider the Lorenz gauge choice
\[ \psi = \partial_\mu A^{a\mu} = \partial_i A_i^a - \partial_0 A_0^a \eqno(16) \]
in eq. (6); with this
\[ \psi + \left\lbrace \psi , G\right\rbrace = \partial_\mu A^{a\mu} + \left\lbrace \partial_i A_i^a - \partial_0 A_0^a, (-D_0^{cd} \mu_2^b)\pi^{c0} + \mu_2^c D_i^{cd} \pi^{di} \right\rbrace  .\eqno(17) \]
From the integral over $\mu_2^a$ in eq. (6), it follows that 
\[ \Delta = \det \left[ \frac{\partial}{\partial t} D_0^{cd} - \frac{\partial}{\partial x^i} D_i^{ab}\right] \eqno(18) \]
\[ \hspace{-1cm}= \det (-\partial^\mu D_\mu^{ab}) \nonumber \]
\[\hspace{3cm} = \int D \overline{c}\, Dc^a \exp i \int dx (\partial^\mu \overline{c}^a) (D_\mu^{ab} c^b).\nonumber \]
We see from eq. (18) that we have recovered the usual Faddeev-Popov determinant associated with the gauge condition of eq. (16), which is exponentiated using the Fermionic ghost fields $c^a$, $\overline{c}^a$.

We also insert the constant
\[ \mathrm{const.} = \int Dk\, e^{i\int dx(-\frac{1}{2\alpha} k^2(x))} \eqno(19) \]
into the path integral of eq. (3). The gauge transformation generated by $-G$ is then performed in the path integral.  We know that the total action and the determinant $\Delta$ are gauge invariant. Upon integrating over $k(x)$, we are then left with 
\[ Z[J] = \int D\Phi\, D\Pi\, D\lambda_{a_{1}}\, \Delta \exp i \int dx \left( \Pi \frac{\partial \Phi}{\partial t} - \mathcal{H}_c - \lambda_{a_{1}}\
\phi_{a_{1}} - \frac{1}{2\alpha} (\psi)^2 + J\Phi\right) \eqno(20) \]
which for Yang-Mills becomes
\[\hspace{-2cm} = \int DA_\mu^a D\pi^{a\mu} \det (-\partial^\mu D_\mu^{ab}) \delta(\pi^{a0})\exp i \int dt\left[ \pi^{a\mu} \frac{\partial A^a_\mu}{\partial t} \right. \eqno(21) \]
\[\left.  - \left(\frac{1}{2} \pi^{ai} \pi^{ai} + \frac{1}{4} F_{ij}^a F_{ij}^a + \pi^{ai} D_i^{ab} A_0^b\right) - \frac{1}{2a} (\partial_\mu A^{a_{\mu}})^2 + J^{a\mu} A_\mu^a\right]. \nonumber \]
In order to convert the argument of the exponential in eq. (21) to being the action $\int dx \mathcal{L}$ we employ the technique outlined in ref. [57-59] and reviewed in appendix C. Following the argument that leads to eq. (C.20), we end up with the usual expression
\[ Z[J] = \int DA_\mu^a \det (-\partial^\mu D_\mu^{ab}) \exp i \int dx \left[ -\frac{1}{4} F_{\mu\nu}^a (A) F^{a\mu\nu}(A)
 - \frac{1}{2\alpha} (\partial \cdot A^a)^2 + J^{a\mu} A_\mu^a\right], \eqno(22) \]
as the integral over $v_i$ in eq. (C.21) is a constant for the Yang-Mills model.

We now turn our attention to spin two fields to illustrate our approach to quantization of gauge models using the path integral.

\section{Spin Two}

The first order form of the Lagrangian for a spin two field is in $d$ dimensions 
\[ \mathcal{L} = h^{\mu\nu} G_{\mu\nu,\lambda}^\lambda +  \eta^{\mu\nu} \left(\frac{1}{d-1} G_{\lambda\mu}^\lambda G_{\sigma\nu}^\sigma -G_{\sigma\mu}^\lambda G_{\lambda\nu}^\sigma\right)\eqno(23) \]
where $h^{\mu\nu} = h^{\mu\nu}$, $G_{\mu\nu}^\lambda = G_{\nu\mu}^\lambda$ and $\eta^{\mu\nu} = \mathrm{diag} (+ + + \ldots -)$.  The gauge transformation 
\[\hspace{-1.6cm}\delta h^{\mu\nu} = \partial^\mu f^\nu + \partial^\nu f^\mu - \eta^{\mu\nu} \partial \cdot f\eqno(24a) \]
\[ \delta G_{\mu\nu}^\lambda = -\partial_{\mu\nu}^2 f^\lambda + \frac{1}{2} \left(\delta_\mu^\lambda \partial_\nu + \delta_\nu^\lambda \partial_\mu\right)\partial \cdot f\eqno(24b) \]
leaves the actions associated with eq. (23) invariant. The generator $G$ associated with this gauge transformation has been discussed in ref. [60]. The equation of motion for $G_{\mu\nu}^\lambda$ results in
\[ \hspace{-1cm} G_{\mu\nu}^\lambda = h_{,\rho}^{\pi\tau} \left[ -\frac{1}{2(d-2)} \eta_{\mu\nu} \eta^{\lambda\rho}\eta_{\pi\tau} + \frac{1}{4} \eta^{\lambda\rho}\left(\eta_{\mu\pi}\eta_{\nu\tau} + \eta_{\mu\tau}
\eta_{\nu\pi}\right) \right. \eqno(25) \]
\[ \left.\hspace{2cm}  -\frac{1}{2(d-2)} \left(\delta_\mu^\rho \delta_\pi^\lambda \eta_{\nu\tau} + 
\delta_\nu^\rho \delta_\pi^\lambda \eta_{\mu\tau} + 
\delta_\mu^\rho \delta_\tau^\lambda \eta_{\nu\pi} +
\delta_\nu^\rho \delta_\pi^\lambda \eta_{\mu\tau} \right)\right] \nonumber\]
which, when $d = 4$, leads to [61]
\[ \mathcal{L} = -\frac{1}{2} \partial_\mu h^{\nu\lambda} \partial^\mu h_{\nu\lambda} + \frac{1}{4} \partial_\mu h^\lambda_{\;\;\;\lambda} \partial^\mu h^\sigma_{\;\;\;\sigma} + \partial _\mu h^{\mu\lambda} \partial^\nu h_{\nu\lambda}\eqno(26) \]
upon being substituted back into eq. (23). The canonical structure of this second order action has been discussed in refs. [62,63]; we will reconsider this using the convenient parameterization of $h^{\mu\nu}$ given by
\[ h^{00} = h, \quad h^{0i} = h^i,\quad h^{ij} = H^{ij} + \delta^{ij}(t + h)\eqno(27)\]
with $H^{ii} = 0$.  In terms of these fields, 
\[ \mathcal{L} = -\frac{1}{2} H^{ij}_{,k} H^{ij}_{,k} + \frac{7}{4} t_{,k} t_{,k} + 2  t_{,k} h_{,k} + h^i_{,j} h^i_{,j} + \frac{1}{2} \dot{H}^{ij} \dot{H}^{ij}\nonumber \]
\[ - \frac{3}{4} \dot{t}^2 + H_{,i}^{ik} H_{,j}^{jk} + 2H^{ij}_{\;\;\;,j} (t + h)_{,i} + 2 \dot{h}^i H^{ij}_{\;\;\;,j} \eqno(28) \]
\[ -h_{,i}^i h_{,j}^j - 2 \dot{h} h_{,i}^i + 2\dot{h}^i(t + h)_{,i}.\nonumber \]
From this, we see that the momenta conjugate to $H^{ij}$, $t$, $h$ and $h^i$ are respectively 
\[ \Pi_{ij} = \dot{H}^{ij}, \quad \tau = - \frac{3}{2} \dot{t}, \quad \pi = -2 h^i_{,i}, \quad \pi_i = 2H_{,j}^{ij} + 2(t + h)_{,i} \eqno(29a-d) \]
with (29c,d) being a pair of primary constraints, $\phi_1$ and $\phi_{1i}$ respectively. The Hamiltonian that follows from eq. (28) is 
\[ \mathcal{H}_c = \frac{1}{2} \Pi_{ij} \Pi_{ij} - \frac{1}{3} \tau^2 + \frac{1}{2} H^{ij}_{\;\;\;,k}H^{ij}_{\;\;\;,k} - \frac{7}{4} t_{,k}t_{,k}\eqno(30) \]
\[-2t_{,k}h_{,k} - h_{,j}^i h_{,j}^i - H_{,i}^{ik} H_{,j}^{jk} + h_{,i}^i h_{,j}^j\nonumber \]
\[-2H_{\;,j}^{ij}(t + h)_{,i} ;\nonumber\]
we find that the primary constraints generate a pair of secondary constraints as with $H_c = \int dx \mathcal{H}_c$
\[ \hspace{-2.4cm}\phi_2 = \left\lbrace \phi_1, H_c \right\rbrace = -2(t_{,ii} + H^{ij}_{\;\;\;,ij}) \eqno(31a)\]
\[ \phi_{2i} = \left\lbrace \phi_{1i}, H_c \right\rbrace = 2
\left( h_{,ij}^j - h_{,jj}^i -\Pi_{ij,j} + \frac{2}{3} \tau_{,i} \right).\eqno(31b)\]
No higher generation of constraints appears, and these constraints are all first class as any pair of them has a vanishing PB with each other. We have 
\[\left\lbrace \phi_2, H_c\right\rbrace = \phi_{2i,i} \qquad 
\left\lbrace \phi_{2i}, H_c\right\rbrace = \phi_{2,i} .\eqno(32a,b)\]
The gauge generator that follows from applying eq. (B.5) is 
\[ G = \int d^3x \left[ (-\dot{\mu}_2 +\mu_{2i,i})\phi_1 +  
(-\dot{\mu}_{2i} +\mu_{2,i}) \phi_{1i} + \mu_2\phi_2 + \mu_{2i} \phi_{2i}\right]; \eqno(33) \]
this generates the gauge transformation of eq. (24a) provided 
$f^\mu = (-\mu_{2i}, -\mu_2)$. 

We will now consider how the path integral can be used to quantize this free spin two field. If we first apply the Faddeev-Popov procedure outlined in Appendix A, then from eq. (A.7) we see that with the most general gauge fixing [64,65], the generating functional is 
\[ Z[J] = \int Dh^{\mu\nu} D\theta^\mu \exp i \int dx \bigg[ \left( - \frac{1}{2} \partial_\mu h^{\mu\lambda}  \partial^\mu h_{\nu\lambda} + \frac{1}{4} \partial_\mu h^\lambda_{\;\;\lambda} \partial^\mu h^\sigma_{\;\;\sigma}\right.\nonumber\]
\[ + \partial_\mu h^{\mu\lambda} \partial^\nu h_{\nu\lambda} - \frac{1}{2\alpha} \left( a\partial_\lambda h^{\lambda\mu} + \partial^\mu h^\lambda_{\,\,\lambda}\right)\nonumber \]
\[ \times (g_{\mu\nu} + \beta\partial_\mu \partial_\nu/\partial^2)(b\partial_\sigma h^{\sigma\nu} + \partial^\nu h^\sigma_{\,\,\sigma} + b\partial^2 \theta^\nu - 2\partial^\nu \partial \cdot \theta)\bigg] \]
\[ \det(g_{\mu\nu} + \beta \partial_\mu\partial_\nu/\partial^2)\det (a \partial^2 \eta^{\mu\nu} - 2\partial^\mu\partial^\nu)\nonumber \]
\[ \det(b\partial^2 \eta^{\mu\nu} - 2\partial^\mu\partial^\nu).\eqno(34) \]
As is discussed in ref. [64], only if $a \neq b$ can one obtain a propagator $D_{\mu\nu , \lambda\sigma}(k)$ for the field $h^{\mu\nu}$ that satisfies the ``transverse-traceless'' (TT) condition
\[ \eta^{\mu\nu} D_{\mu\nu , \lambda\sigma}(k) = 0 = k^\mu 
 D_{\mu\nu , \lambda\sigma}(k).\eqno(35) \]
In the limit that there is no Nielsen-Kallosh contribution and there is but a single gauge condition $\partial_\lambda h^{\lambda\mu} = 0$, eq. (34) reduces to the standard form 
\[ Z[J] = \int Dh^{\mu\nu} \exp i \int dx \left[ -\frac{1}{2} \partial_\mu h^{\mu\lambda} \partial^\mu h_{\nu\lambda} + 
\frac{1}{4} \partial_\mu h^{\lambda}_{\;\;\lambda} \partial^\mu h^\sigma_{\;\;\sigma} \right. \nonumber \]
\[ \left. + \partial_\mu h^{\mu\lambda} \partial^\nu h_{\nu\lambda} - 
\frac{1}{2\alpha} (\partial_\lambda h^{\lambda \mu})(\partial^\sigma h_{\lambda\mu})\right] \det (\partial^2\eta^{\mu\nu}). \eqno(36) \]

It is of interest to apply Faddeev's general formula of eq. (4) to obtain the generating functional for the free spin two field.  We will begin by accompanying the first class constraints $\phi_1$, $\phi_{1i}$, $\phi_2$, $\phi_{2i}$ of eqs. (29c, 29d, 31a, 31b) respectively by the gauge conditions 
\[ \chi_1 = h\;, \quad \chi_{1i} = h^i \;, \quad 
\chi_2 = \frac{1}{2} (\tau - \pi)\;, \quad  
\chi_{2i} = \frac{1}{2} H^{ij}_{\;\;,j} \eqno(37a-d) \]
so that
\[\hspace{-2.5cm} \left\lbrace \chi_1, \phi_1 \right\rbrace = 1 \eqno(38a) \]
\[\hspace{-2.5cm} \left\lbrace \chi_{1i}, \phi_{1j} \right\rbrace =  \delta_{ij} \eqno(38b) \]
\[\hspace{-2.5cm} \left\lbrace \chi_{2}, \phi_{2} \right\rbrace = \nabla^2 \eqno(38c) \]
\[ \left\lbrace \chi_{2i}, \phi_{2j} \right\rbrace = \frac{1}{2}\nabla^2
\delta_{ij} + \frac{1}{6} \partial_i \partial_j, \eqno(38d) \]
with all other PB between gauge conditions and constraints vanishing. The contribution to eq. (4) coming from the gauge conditions and first class constraints reduces to 
\[\hspace{-2cm}\mathop{\Pi}_a \delta(\phi_a) \delta(\chi_a) = \delta(h) \delta(h^i) \delta(\frac{1}{2} \tau) \delta(\frac{1}{2} H^{ij}_{\;\;,j}) 
 \delta(\pi) \delta(\tau_i - 2t_{,i}) \delta(-2t_{,ii}) \delta(-2\pi_{ij,j} \pi_{ij,j}.)\eqno(39) \]
With the constraints imposed by these delta functions, we end up with
\[ \int dx \left( \Pi(x) \frac{\partial\Phi (x)}{\partial t} - \mathcal{H}_c \right) = \int dx \left[ \Pi_{ij} \dot{H}^{ij} - \left( \frac{1}{2} \Pi_{ij} \Pi_{ij} + \frac{1}{2} H^{ij}_{\;\;,k} H^{ij}_{\;\;,k}\right)\right]\eqno(40) \]
and so the generating functional of eq. (4) for the free 
spin two field is 
\[\hspace{-6cm} Z[J] = \int D H^{ij} D\Pi_{ij} \delta(H^{ij}_{\;\;,j})\det(\nabla^2)\det \left(\frac{1}{2} \nabla^2 \delta_{ij} + \frac{1}{6} \partial_i\partial_j\right) \eqno(41) \]
\[ \exp i \int dx \left[ \Pi_{ij} \dot{H}^{ij} - \frac{1}{2} (\Pi_{ij} \Pi_{ij} + H^{ij}_{\;\;,k} H^{ij}_{\;\;,k}  + J_{\mu\nu} h^{\mu\nu}\right] \nonumber \]
when using the approach of Faddeev in ref. [10]. This demonstrates that only the transverse, traceless part of the spatial components of $h^{\mu\nu}$ are dynamical.  A non-covariant gauge fixing can be used in conjunction with the Faddeev-Popov approach to the path integral to yield a propagator that only involves the transverse, traceless parts of the spatial components of $h^{\mu\nu}$ [66,67].  However, it is not clear if the path integral of eq. (41) is equivalent in all respects to that of eq. (36).

We now turn to applying eq. (6) to obtain an expression for the path integral of eq. (3) that involves covariant gauge fixing. This is more complicated than the case of Yang-Mills theory, as there now are two primary first class constraints rather than one. We now consider the covariant gauge fixing
\[ \psi^\nu = \partial_\mu h^{\mu\nu}\eqno(42) \]
which for $\nu = 0$ becomes 
\[ \psi^0 = \dot{h} + h_{,i}^i \eqno(43a) \]
and for $\nu = i$ becomes 
\[ \psi^i = \dot{h}^i + H^{ij}_{\;\;,j} + (t + h)_{,i} \eqno(43b) \]
with the parameterization of eq. (27).  When we evaluate the PB appearing in eq. (6), we must keep in mind that 
\[ \left\lbrace H^{ij}, \Pi_{k\ell}\right\rbrace = \frac{1}{2} \left( \delta_k^i \delta_\ell^j + \delta_\ell^i \delta_k^j \right) - \frac{1}{3} (\delta^{ij} \delta_{k\ell}) \eqno(44) \]
since $H^{ii} = 0$. With $G$ being given by eq. (33) it follows that 
\[ \left\lbrace \psi^0, G\right\rbrace = \mu_2 \left(\nabla^2 - \frac{\partial^2}{\partial t^2} \right) \eqno(45a)\]
\[ \left\lbrace \psi^i, G\right\rbrace = \mu_{2i} \left(\nabla^2 - \frac{\partial^2}{\partial t^2} \right). \eqno(45b)\]
By eq. (45), the functional determinant appearing in eq. (6) is consequently 
\[\Delta = \det (\partial^2 \eta^{\mu\nu} )\eqno(46) \]
as in eq. (36). This is the usual Faddeev-Popov determinant.

When converting the phase space path integral of eq. (20) to the configuration space path integral, we use eqs. (C.20, C.21) with the identifications
\[ q_i^\prime \rightarrow (H^{ij}, t) \eqno(47a) \]
\[ q_i^{\prime\prime} \rightarrow ( h, h^i). \eqno(47b) \]
The various contributions to $\Lambda_r$ that occur can be seen to be 
\[ L(q_i, v_i +\dot{q}^\prime_i, \dot{q}_i^{\prime\prime}) - 
L(q_i, \dot{q}^\prime_i, \dot{q}_i^{\prime\prime}) - v^i \frac{\partial}{\partial v^i}  L(q_i, v_i +\dot{q}^\prime_i, \dot{q}_i^{\prime\prime}) = -\frac{1}{2} v^{ij} v^{ij} + \frac{3}{4} v^2\;.\eqno(48) \]
We then find that 
\[ A_r \left(q_i, \frac{\partial L}{\partial v_i} \left( q_i, v_i + 
\dot{q}^\prime_i, \dot{q}^{\prime\prime}\right) \right) = \det \left( 
\begin{array}{cc}
-\frac{3}{2} & 0 \\
0 & \frac{1}{2}(\delta^{ik} \delta^{j\ell} + \delta^{i\ell} \delta^{jk}) - \frac{1}{3} \delta^{ij} \delta^{k\ell} 
\end{array}
\right) \eqno(49) \]
which is a constant.  Finally, we must consider $\delta(\phi_A (q_i, \frac{\partial}{\partial v_i} L(q_i, v_i + \dot{q}^\prime_i, \dot{q}^{\prime\prime}), g^i ))$ which for the two primary first class constraints associated with our free spin two field becomes 
\[ \delta(\phi_1) = \delta(0)\;; \quad \delta(\phi_{1i}) = \delta(0)\;;  \eqno(50a-b)\]

The delta functions of eqs. (50a,b) can be absorbed into the normalization of the path integral. We thus see that this approach to the path integral in phase space leads to the same path integral (eq. (36)) as was obtained by using the Faddeev-Popov procedure in configuration space.  It is also apparent that just as one can generalize the original Faddeev-Popov procedure (see eq. (A.7)), so also eq. (6) can be generalized so as to obtain eq. (34) from the phase space path integral.  We also note that only by having included the primary first class constraints $\phi_1$, $\phi_{1i}$ and not the secondary first class constaints $\phi_2$, $\phi_{2i}$ in the total Hamiltonian $\mathcal{H}_T$ appearing in eq. (3) has it been possible to recover eq. (36) from the phase space path integral for the spin two field.

Although we have been dealing with the free spin two field in four dimensions, it is apparent that we should extend these considerations to applying the path integral to quantizing general relativity. A standard approach is to take the Einstein-Hilbert (EH) Lagrangian
\[ \mathcal{L}_{EH} = \sqrt{-g}\; g^{\mu\nu} R_{\mu\nu} (g) \eqno(51)\]
and expand the metric $g_{\mu\nu}$ about the flat metric $\eta_{\mu\nu}$, 
\[ g_{\mu\nu} = \eta_{\mu\nu} + \gamma_{\mu\nu} \eqno(52) \]
so that we obtain an (infinite) power series in $\gamma_{\mu\nu}$ for $\mathcal{L}_{EH}$.  The lowest order (quadratic) term in $\gamma_{\mu\nu}$ is that of eq. (26) and the gauge transformation of eq. (24a) is the lowest order form of the renormalizable diffeomorphism invariance that is present in eq. (51) [61].  The Faddeev-Popov quantization procedure, in conjunction with dimensional regularization, was used in refs. [66, 68, 69] to demonstrate that up to two loop order, $\mathcal{L}_{EH}$ is not renormalizable.  (Operator regularization was applied to this problem in ref. [70] thereby avoiding the occurance of explicit divergences.)  The canonical structure of the EH action was not considered in this approach; it is not readily apparent if in fact for this model an analysis of its canonical structure would lead to a path integral quantization that is consistent with the Faddeev-Popov approach. This problem is especially noted in the second paper of ref. [6].  The canonical structure of this second order Lagrangian $\mathcal{L}_{EH}$ has been examined in a number of papers [71-75]; the last two in particular show that the canonical Hamiltonian is a linear combination of first class constraints that can be used to obtain the generator of the diffeomorphism transformation in four dimensions. (This is unlike the first class constraints appearing in the analysis of ref. [73] which appear to lead to a generator of the diffeomorphism only in the spatial dimensions [55, 76].)  Direct application of this canonical structure to the path integrals of either eqs. (4) or (C.20) has not been effected, though the BFV approach is discussed at length in ref. [13].

Since $\mathcal{L}_{EH}$ is non-polynomial in $\gamma_{\mu\nu}$, one might consider just using the first order ``Palatini'' form of the gravitational action 
\[ \mathcal{L}_{I\!\!P} = \sqrt{-g}\; g^{\mu\nu} R_{\mu\nu} (\Gamma) \eqno(53) \]
(though this form of the action actually originated [77] with Einstein [78].)  In eq. (53), provided we are in $d > 2$ dimensions, the equation of motion for the affine connection $\Gamma_{\mu\nu}^\lambda$ can be solved to   
yield 
\[ \Gamma_{\mu\nu}^\lambda = \frac{1}{2} g^{\lambda\sigma} \left( g_{\sigma\mu ,\nu} + g_{\sigma\nu ,\mu} - g_{\mu\nu , \sigma}\right) \eqno(54) \]
which, when substituted back into eq. (53), leads to eq. (51).

The advantage to using eq. (53) is that if the independent fields are taken to be $h^{\mu\nu} = \sqrt{-g}\; g^{\mu\nu}$ and $G_{\mu\nu}^\lambda = \Gamma_{\mu\nu}^\lambda - \frac{1}{2}(\delta_\mu^\lambda \Gamma^\sigma_{\sigma\nu} + \delta_\nu^\lambda \Gamma_{\sigma\mu}^\sigma )$ 
then we have
\[ \mathcal{L}_{I\!\!P} = h^{\mu\nu} \left[ G_{\mu\nu ,\lambda}^\lambda + \frac{1}{d-1}  G_{\lambda\mu}^\lambda G_{\sigma\nu}^\sigma - G_{\sigma\mu}^\lambda G_{\lambda\nu}^\sigma \right] \eqno(55) \]
so that the interaction term is at most cubic in $h$ and $G$ [79, 80].

If one were to simply break the infinitesmal gauge invariance associated with diffeomorphism invariance present in eq. (55) 
\[\hspace{-1.5cm}\delta h^{\mu\nu} = h^{\mu\lambda}\partial_\lambda \theta^\nu + 
h^{\nu\lambda}\partial_\lambda \theta^\mu - \partial_\lambda (h^{\mu\nu}\theta^\lambda )\eqno(56a) \]
\[\delta G_{\mu\nu}^\lambda = - \partial_{\mu\nu}^2 \theta^\lambda + \frac{1}{2} (\delta_\mu^\lambda \partial_\nu + \delta_\nu^\lambda \partial_\mu) \partial \cdot \theta - \theta \cdot \partial G_{\mu\nu}^\lambda \eqno(56b) \]
\[\hspace{1cm} + G_{\mu\nu}^\rho \partial_\rho\theta^\lambda - (G_{\mu\rho}^\lambda \partial_\nu + G_{\nu\rho}^\lambda \partial_\mu)\theta^\rho\nonumber \]
by some choice of gauge, it is unfortunately still not possible to directly apply the Faddeev-Popov procedure (as outlined in Appendix A) to the Lagrangian of eq. (55), as it is not possible to find a suitable gauge choice that allows for one to invert those terms in the action that are bilinear in the fields. However, if in eq. (55), one were to expand $h^{\mu\nu}$ about a flat background so that much like in eq. (52) 
\[ h^{\mu\nu} \rightarrow \eta^{\mu\nu} + h^{\mu\nu} \eqno(57) \]
then the bilinear terms in eq. (57) yield $\mathcal{L}_{I\!\!P}^{(2)}$ which is simply the first order form of the spin two action of eq. (23); the gauge transformations of eq. (24) are the lowest order form of these of eq. (56). One now finds that with the gauge fixing Lagrangian
\[ \mathcal{L}_{gf} = - \frac{1}{2} (\partial_\mu h^{\mu\nu})^2\eqno(58) \]
the bilinear terms in the action is 
\[ \mathcal{L}_p^{(2)} + \mathcal{L}_{gf} = \frac{1}{2} (h^{\mu\nu}, G_{\mu\nu}^\rho) \left( \begin{array}{cc}
\frac{1}{4}(\partial_\mu\partial_\lambda\eta_{\nu\sigma} + \partial_\nu\partial_\lambda\eta_{\mu\sigma} & \Delta_{\mu\nu}^{\gamma\delta} \partial_\kappa \\
 \hspace{1cm}+ \partial_\mu\partial_\sigma\eta_{\nu\lambda} + \partial_\nu\partial_\sigma\eta_{\mu\lambda}) & \\
 \\
-\Delta_{\lambda\sigma}^{\alpha\beta} \partial_\rho & D_{\rho\;\;\;\;\;\kappa}^{\mu\nu \;\;\gamma\delta} 
\end{array} \right)\left(\begin{array}{c}
h^{\lambda\sigma} \\
G_{\gamma\delta}^\kappa \end{array}\right) \eqno(59) \]
where 
\[ \Delta_{\mu\nu}^{\gamma\delta} = \frac{1}{2} \left( \delta_\mu^\gamma \delta_\nu^\delta +\delta_\nu^\gamma \delta_\mu^\delta \right) \eqno(60) \]
and 
\[ \hspace{-.7cm}D_{\rho\;\;\;\;\;\kappa}^{\mu\nu \;\;\gamma\delta}  = \frac{1}{2(d-1)} \left[ \delta_\rho^\mu \delta_\kappa^\gamma \eta^{\nu\delta} + 
\delta_\rho^\mu \delta_\kappa^\delta \eta^{\nu\gamma} + 
\delta_\rho^\nu \delta_\kappa^\gamma \eta^{\mu\delta} + 
\delta_\rho^\nu \delta_\kappa^\delta \eta^{\mu\gamma} \right]\nonumber \]
\[ -\frac{1}{2} \left[ \delta_\kappa^\mu \delta_\rho^\gamma \eta^{\nu\delta} + 
\delta_\kappa^\mu \delta_\rho^\delta \eta^{\nu\gamma} + 
\delta_\kappa^\nu \delta_\rho^\gamma \eta^{\mu\delta} + 
\delta_\kappa^\nu \delta_\rho^\delta \eta^{\mu\gamma} \right].\eqno(61)\]
The matrix $\mathbf{M}$ in eq. (59) has an inverse that can be found using the standard relation
\[ \left( \begin{array}{cc}
A & B \\
C & D \end{array} \right)^{-1} = 
\left( \begin{array}{cc}
(A-BD^{-1}C)^{-1} & -(A-BD^{-1}C)^{-1}BD^{-1} \\
-D^{-1}C(A-BD^{-1}C)^{-1} & D^{-1}+D^{-1}C(A - BD^{-1}C)^{-1}BD^{-1}
\end{array} \right).\eqno(62) \]
We find that for $\mathbf{M}$,
\[\hspace{-4cm} \left[ (A-BD^{-1} C)^{-1}\right]^{\mu\nu ,\lambda\sigma} = \left(\eta^{\mu\lambda} \eta^{\nu\sigma} + \eta^{\nu\lambda} \eta^{\mu\sigma} - \eta^{\mu\nu}\eta^{\lambda\sigma}\right) \frac{1}{\partial^2} \eqno(63a) \]
\[ \left[ D^{-1} \right]^{\lambda\;\;\;\;\;\rho}_{\mu\nu\; ,\;\pi\tau} = -\frac{1}{2(d-2)} \eta_{\mu\nu} \eta^{\lambda\rho}\eta_{\pi\tau} - \frac{1}{4} \left( \delta_\mu^\rho \delta_\pi^\lambda \eta_{\nu\tau} + 
\delta_\mu^\rho \delta_\tau^\lambda \eta_{\nu\pi}\right.\nonumber \eqno(63b) \]
\[\hspace{3.4cm}\left. + \delta_\nu^\rho \delta_\pi^\lambda \eta_{\mu\tau} +
\delta_\nu^\rho \delta_\tau^\lambda \eta_{\mu\pi} \right) + \frac{1}{4} \eta^{\lambda\rho} (\eta_{\mu\pi} \eta_{\nu\tau} + \eta_{\mu\tau}\eta_{\nu\pi}) \nonumber \]
\[ \hspace{-3.5cm} \left[  B\right]^{\;\;\;\;\;\gamma\delta}_{\mu\nu\; ,\;\kappa} = \Delta_{\mu\nu}^{\gamma\delta} \partial_\kappa = - \left[ C \right]^{\gamma\delta}_{\kappa , \mu\nu} . \eqno(63c) \]
From eqs. (62-63) one can find the propagators $\langle hh\rangle$, $\langle GG\rangle$ as well as the mixed propagators $\langle hG\rangle$, $\langle G,h\rangle$. The $h-G-G$ term in eq. (55) fixes a three point vertex when using the Faddeev-Popov procedure; in addition there is the contribution coming from the Faddeev-Popov determinant arising from eqs. (56a, 57, 58)
\[ \exp i \int dx \left[ \overline{c}_\mu \partial^2 c^\mu + \overline{c}_\mu \left( \partial_\lambda (h^{\lambda\sigma} \partial_\sigma c^\mu) \right.\right. \eqno(64) \]
\[\hspace{3cm} \left. \left.  + \partial_\lambda (h^{\mu\sigma} \partial_\sigma c^\lambda ) - \partial_\lambda \partial_\sigma (h^{\lambda\mu} c^\sigma) \right)\right] \nonumber \]
where $c^\mu$, $\overline{c}^{\,\mu}$ are Fermionic vector ghost fields.

However, it is not clear if the Faddeev-Popov quantization procedure is equivalent to the path integral of eq. (5) which follows from canonical quantization for the first order Palatini Lagrangian of eq. (53).  An argument has been given [80] that the path integral derived from canonical quantization is equivalent for this model to the one following from the manifestly covariant Faddeev-Popov approach. (The two forms of the path integral associated with the second order EH action are related in ref. [13].)  However, the analysis of the constraints associated with $\mathcal{L}_{I\!\!P}$ appearing in refs. [80, 81] neglects considering the second class constraints that are present and does not find any tertiary first class constraints. Rather, in these references, equations of motion that are independent of time derivatives are used to eliminate fields from the action and it is only at this stage is the action analyzed using the Dirac constraint formalism.  In doing this, one does not encounter tertiary first class constraints, which means that the generator of gauge transformations cannot produce the second derivatives of gauge functions present in eq. (56b), as can be seen from eq. (B.5).  If one consistently applies the Dirac constraint formalism at the outset to the Palatini Lagrangian of eq. (53), then both second class constraints and tertiary first class constraints occur [60, 82].

It is not clear if these additional constraints that occur in the treatment of the Palatini action that appear in the treatment of refs. [60, 82] (but are absent in the treatments of refs. [80, 81]) modify the path integral that follows from the canonical formalism so that it is no longer equivalent to the path integral that follows from the Faddeev-Popov approach. In particular, the second class constraints $\theta_a$ that arise in the treatment of refs. [60, 82] contribute in a non-trivial way to the measure of eq. (5) as in this instance the PB $\left\lbrace \theta_a, \theta_b \right\rbrace$ of the second class constraints is field dependent.

There is a precedent for the presence of second class constraints leading to a non-trivial contribution to the measure of the path integral.  The model of ref. [33] (which involves a vector gauge field $W_\mu^a$ coupled to an antisymmetric tensor field $\phi_{\mu\nu}^a$ that possesses a pseudoscalar mass term) has been quantized in refs. [34, 35, 36] using the Faddeev-Popov path integral.  The ensuing calculation of the two point functions in refs. [35, 36] leads to divergences whose structure is inconsistent with the internal structure of the theory.  It has been proposed that this inconsistency is a consequence of having ignored the contribution of non-trivial second class constraints to the measure of the path integral [37].  (We note though that the PB $\left\lbrace \theta_a, \theta_b \right\rbrace $ of these second class constraints leads to a contribution to the measure in eq. (5) that is not manifestly covariant.) It is quite possible that ignoring the second class constraints associated with the Palatini Lagrangian of eq. (53) results in the path integral derived from the canonical structure of the theory being inequivalent to the path integral that follows from the Faddeev-Popov approach.

There have been a number of proposals for ways of reconsidering models with second class constraints so that from an alternate view point, one has only first class constraints. One idea, is to split the second class constraints into two portions with half of them being taken to be first class constraints and the other half the associated gauge conditions; the Hamiltonian is then altered so that when the ``gauge conditions'' are imposed the original Hamiltonian is recovered [47-50].  Another approach (BFT) is to supplement the number of dynamical variables in phase space so that when the second class constraints and Hamiltonian are assigned appropriate dependence on these new variables, all constraints are first  class and when the associated gauge condition is taken to be the vanishing of these new variables, the original model in phase space is recovered [38-46].

Neither of these approaches is easy to implement unless the model in question is particularly simple.  Furthermore, it is not clear if the modified model in which there are no second class constraints can be related to a Lagrangian in an associated configuration space that is covariant.  An exception occurs when the BFT approach involves an introduction of new variables in phase space that is equivalent to Stueckelberg's introduction of a new field into the configuration space Lagrangian in order to restore a gauge symmetry that has been explicitly broken [44]. (This happens in the case of a Proca vector fields [42, 43].)  If this is possible the Lagrangian can be made explicitly covariant.

Other approaches to dealing with second class constraints in the path integral are given in refs. [51, 52].

Currently, we are examining in more detail how the Palatini action can be quantized by using the phase space path integral in conjunction with eq. (6) and by ``integrating out'' the second class constraints in the theory.

\section{The First Order Einstein-Hilbert Action in Two Dimensions}

The comments on the Palatini Lagrangian of eq. (55) appearing in the preceding section, especially those concerning second class constraints, are pertinent only if $d > 2$.  When $d = 2$, a number of interesting features occur.  First of all, the first and second order forms of the Einstein-Hilbert action are no longer equivalent if $d = 2$.  Secondly, only first class constraints occur when making a canonical analysis of this action; there are no second class constraints.  In addition, the first class constraints lead to a gauge generator that is associated with a gauge transformation that is distinct from the diffeomorphism transformation of eq. (56).  These features are explored in detail in refs. [83-86]; we will now summarize them. 

If, when $d = 2$, we define
\[ \pi = - G_{00}^0 \qquad \pi_1 = -2G_{01}^0 \qquad \pi_{11} = -G_{11}^0 \eqno(65a-c) \]
\[ \xi = - G_{00}^1 \qquad \xi = 2G_{01}^1 \qquad \xi_1 = G_{11}^1 \eqno(66a-c) \]
then the action of eq. (55) leads to the canonical Hamiltonian
\[  \mathcal{H}_c = \xi^1 \phi_1 + \xi\phi + \xi_1 \phi^1 \eqno(67) \]
where 
\[ \phi_1 = h_{,1} - h\pi_1 -2h^1\pi_{11} \eqno(68a) \]
\[ \phi = h_{,1}^1 + h\pi - h^{11}\pi_{11} \eqno(68b) \]
\[ \phi^1 = h_{,1}^{11} + 2h^1\pi  + h^{11}\pi_1 .\eqno(68c) \]
A set of six second class constraints fix $(\pi , \pi_1, \pi_{11} )$ to be the momenta conjugate to $(h , h^1, h^{11})$ respectively. Three primary first class constraints are that the momenta conjugate to $(\xi^1, \xi, \xi_1)$ all vanish; by eq. (67) this leads to the secondary first class constraints $\phi_1 = \phi = \phi^{1} = 0$ whose PB algebra is 
\[ \left\lbrace \phi_1 , \phi^1 \right\rbrace = 2\phi\; ,\quad 
\left\lbrace \phi , \phi^1 \right\rbrace = \phi^1\; , \quad 
\left\lbrace \phi_1, \phi \right\rbrace = \phi_1 .\eqno(69a-c) \]
The generator of the gauge transformation that follows from these six first class constraints leads to the gauge transformation 
\[ \delta_A h^{\mu\nu} = - (\epsilon^{\mu\rho} h^{\nu\sigma} + \epsilon^{\nu\rho} h^{\mu\sigma} ) \lambda_{\rho\sigma}^A \eqno(70a) \]
\[ \delta_A G_{\mu\nu}^\lambda = - \epsilon^{\lambda\sigma} \lambda^A_{\mu\nu ,\sigma} - \epsilon^{\rho\sigma} (G_{\mu\rho}^\lambda \lambda_{\nu\sigma}^A + G_{\nu\rho}^\lambda \lambda_{\mu\sigma}^A)\eqno(70b)\]
where $\epsilon^{01} = -\epsilon^{10} = 1$ and $\lambda_{\mu\nu}^A$ is a symmetric gauge function. Eq. (70) is distinct from the manifest diffeomorphism invariance present in the action of eq. (53).  This transformation satisfies the algebra
\[ (\delta_A \delta_B - \delta_B \delta_A ) = \delta_C \eqno(71) \]
where
\[ \lambda_{\mu\nu}^C = -\epsilon^{\alpha\beta} (\lambda_{\mu\alpha}^A \lambda_{\nu\beta}^B + \lambda_{\nu\alpha}^A \lambda_{\mu\beta}^B ). \eqno(72) \]

In applying the Faddeev-Popov quantization procedure a convenient gauge choice is [87]
\[ \epsilon_{\lambda\sigma} G_{\mu\nu}^{\lambda ,\sigma} = 0.\eqno(73) \]
If this is to be incorporated into the action by using a Nakanishi-Lautrup field $N^{\alpha\beta}$, and if we employ symmetric Fermionic ghost fields $\overline{\zeta}^{\,\alpha\beta}$, $\zeta_{\alpha\beta}$, then by eqs. (70b) and (73) we have the effective action 
\[ S_{eff} = \int d^2x \left\lbrace h^{\mu\nu} (G_{\mu\nu ,\lambda}^\lambda + G_{\lambda\mu}^\lambda G_{\sigma\nu}^\sigma - G_{\sigma\mu}^\lambda G_{\lambda\nu}^\sigma )\right.\eqno(74) \]
\[- N^{\alpha\beta} G_{\alpha\beta}^{\lambda ,\sigma} \epsilon_{\lambda\sigma} + \frac{\alpha}{2} N^{\alpha\beta} N_{\alpha\beta}\nonumber \]
\[\hspace{2cm}+ \overline{\zeta}^{\,\lambda\sigma} \epsilon_{\mu\nu} \left[ -\epsilon^{\mu\rho} \zeta_{\lambda\sigma ,\rho} - \epsilon^{\pi\tau} (G^\mu_{\lambda\pi} \zeta_{\tau\sigma} \right. \nonumber \]
\[ \hspace{2cm}\left.\left. + G_{\sigma\pi}^\mu \zeta_{\tau\lambda})\right]^{,\nu}\right\rbrace .\nonumber \]
It turns out that all perturbative radiative corrections to this action vanish [87], which is consistent with there being no dynamical degrees of freedom in the classical action.

If $\Lambda$ is a Grassmann constant, then $S_{eff}$ in eq. (74) is invariant under the BRST transformation (with the Grassmann constant $\Lambda$) [28, 29]
\[ \delta G_{\mu\nu}^\lambda = -\epsilon^{\lambda\rho} \zeta_{\mu\nu ,\rho} \Lambda - \epsilon^{\rho\sigma} (G_{\mu\rho}^\lambda \zeta_{\nu\sigma} + G_{\nu\rho}^\lambda \zeta_{\mu\sigma})\Lambda \eqno(75a) \]
\[\hspace{-2.2cm}\delta h^{\mu\nu} = - (\epsilon^{\mu\rho} h^{\nu\sigma} + \epsilon^{\nu\rho} h^{\mu\sigma}) \zeta_{\rho\sigma} \Lambda \eqno(75b) \]
\[\hspace{-4cm} \delta \zeta_{\alpha\beta} = -\epsilon^{\pi\tau} \zeta_{\alpha\pi} \zeta_{\tau\beta} \Lambda \eqno(75c) \]
\[\hspace{-6.2cm} \delta N^{\alpha\beta} = 0 \eqno(75d) \]
\[\hspace{-4.8cm}\delta \overline{\zeta}^{\alpha\beta} = -N^{\alpha\beta} \Lambda .\eqno(75e) \]

If we were to employ the path integral of eq. (4) in conjunction with this model, it is necessary to choose a gauge condition that is matched with each of the first class constraints.  Gauge conditions appropriate to the three primary first class constraints are
\[ \xi_1 = \xi = \xi^1 = 0; \eqno(76a-c) \]
associated with the three secondary first class constraints $\phi_A \equiv (\phi_1, \phi, \phi^1)$ we select the gauge conditions
\[ \pi_{,1} = \pi_{1,1} = \pi_{11,1} = 0 \eqno(77a-c) \]
respectively. If the constraint $\phi_A$ is associated with a gauge condition $\overline{\phi}_A$, then 
\[ \det \left\lbrace \phi_A, \overline{\phi}_B \right\rbrace = \det \left( \begin{array}{ccc}
-\partial_1^2 & -\pi\partial_1 & \pi_{11}\partial_1 \\
2\pi_1\partial_1 & -\partial_1^2 + \pi_1\partial_1 & 0 \\
-2\pi\partial_1 & 0 & -\partial_1^2 - \pi_1\partial_1 \end{array}
\right) . \eqno(78) \]
The path integral of eq. (4) then becomes
\[\hspace{-4cm} Z[J_{\mu\nu}, J_\lambda^{\mu\nu} ] = \int Dh\,Dh^1\, Dh^{11}\; 
 D\pi\,D\pi_1\, D\pi_{11} \; D\xi^1\,D\xi\, D\xi_1 \nonumber \]
\[ \delta(\xi^1)\delta(\xi)\delta(\xi_1)\delta(\phi_1)\delta(\phi)\delta(\phi^1)
 \delta(\overline{\phi}_1)\delta(\overline{\phi})\delta(\overline{\phi}^1)
 \nonumber \]
\[\hspace{5cm}
 \det\left\lbrace \phi_A, \overline{\phi}_B \right\rbrace \exp i\int d^2x (\pi h_{,0} + \pi_1h^1_{,0} + \pi_{11} h_{,0}^{11} - \xi^1\phi_1 - \xi\phi - \xi_1\phi^1 \nonumber \]
\[ \hspace{3cm} + J_{\mu\nu} h^{\mu\nu} + J_\lambda^{\mu\nu} G_{\mu\nu}^\lambda ).\eqno(79) \]
Exponentiating $\delta(\phi_A)$ by use of Lagrange multipliers reduces eq. (79) to 
\[\hspace{-3cm} Z[J_{\mu\nu}, J_\lambda^{\mu\nu} ] = \int Dh\,Dh^1\, Dh^{11}\; 
 D\pi\,D\pi_1\, D\pi_{11} \; D\lambda^1\,D\lambda\, D\lambda_1 \nonumber \]
\[\hspace{-2cm} \det\left\lbrace \phi_A, \overline{\phi}_B \right\rbrace 
 \delta(\pi_{,1}) \delta( \pi_{1,1}) \delta(\pi_{11,1}) \nonumber \]
\[\hspace{2.5cm} \exp i\int d^2x (\pi h_{,0} + \pi_1h^1_{,0} + \pi_{11} h_{,0}^{11} - \lambda^1\phi_1 - \lambda\phi - \lambda_1\phi^1 \nonumber \]
\[ + J_{\mu\nu} h^{\mu\nu} + J_\lambda^{\mu\nu} G_{\mu\nu}^\lambda ).\eqno(80) \]
It is not clear how to relate the path integrals of eqs. (74) and (80). (That is, it is not clear if the path integral derived from canonical quantization is identical to the one that follows from the Faddeev-Popov procedure when considering this particular model.)

If now we were to consider the path integral appropriate for quantizing this model by using eq. (6) in conjunction with eq. (3), we first need to find the generator of the gauge transformation of eq. (70).  If $(\xi^1, \xi, \xi_1)$ are conjugate to the canonical momenta $(\Pi_1, \Pi, \Pi^1)$ respectively, then the generator is of the form given by eq. (B.2) [83-86]
\[ G = \int dx^1 \left[ (\mu^1 \Pi_1 + \mu\Pi + \mu_1\Pi^1) + (\nu^1 \phi_1 + \nu\phi + \nu_1\phi^1)\right]. \eqno(81) \]
With $\mathcal{H}_c$ given by eq. (67), it follows from eq. (B.5) that 
\[ G = \int dx^1 \left[ (\dot{\nu}^1 + \nu^1\xi - \nu\xi^1) \Pi_1 + (\dot{\nu} + 2\nu^1\xi_1 - 2\nu_1\xi^1)\Pi \right. \eqno(82) \]
\[ \left. + (\dot{\nu}_1 + \nu\xi_1 - \nu_1\xi)\Pi^1 + \nu^1\phi_1 + \nu\phi + \nu_1\phi^1 \right].\nonumber \]
With this generator, it is now straight forward to use eq. (6) and then the general procedure outlined in appendix C to obtain the same result as given in eq.  (74) (which has been shown to follow from the Faddeev-Popov procedure) provided we use the gauge condition of eq. (73) for $\psi$.

\section{Discussion}

We have described how the Faddeev-Popov approach to handling a path integral that involves gauge equivalent fields in configuration space can be adapted to deal with a path integral that involves gauge equivalent fields in phase space.  This approach has the novel feature of permitting us to choose manifestly covariant gauge fixing conditions when dealing with path integrals in phase space.  When using the approach of refs. [57-59] outlined in appendix C for converting these phase space path integrals to configuration space path integrals, we have found that for Yang-Mills theory, for a spin two theory in four dimensions and for the first order Einstein-Hilbert (Palatini) action in two dimensions, the Faddeev-Popov path integral is recovered. The implications of this on quantizing both the first and second order forms of the Einstein-Hilbert action in four dimensions should be worked out.  This would be important, as only the phase space path integral is directly related to canonical quantization.

The problem of dealing with the path integral quantization of models which contain non-trivial second class constraints $\theta_a$ has been noted. The contribution of the factor of $\det^{1/2} \left\lbrace \theta_a, \theta_b \right\rbrace$ in eq. (5) to the measure of the path integral is often just a constant, but if the second class constraints are such that the PB appearing here is field dependent, it is not clear how it can be rendered manifestly covariant.  This problem has been sketched in more detail in ref. [37], where a model involving a vector gauge field and an antisymmetric tensor has been considered.  Various options for converting second class constraints into first class constraints were considered in ref. [37] for this model, but none seem to remove this problem. The second class constraints present in the Palatini action [60, 82] present similar difficulties. 

In our discussion of implementing eq. (6) into the phase space path integral, we have assumed that in each generation there is the same number of first class constraints, so that by eq. (B.5) $\mu_{a_{i}}(i = 1 \ldots N-1)$ are all fixed in terms of $\mu_{a_{N}}(t)$ where $a = 1 \ldots n$ in each generation. Consequently $G$ is fixed in terms of $\mu_{a_{N}}(t)$, and in eq. (6) the integral over $\mu_{a_{N}}(t)$ is an integration over all points in a ``gauge orbit'' in phase space. However, there are models in which the number of first class constraints in each generation is not the same.  For example [88], if the Lagrangian of eq. (55) when $d = 2$ is supplemented by a contribution coming from a scalar field $f^a (a = 1 \ldots M)$
\[ \mathcal{L}_f = \frac{1}{2} h^{\mu\nu} f_{,\mu}^a f_{,\nu}^a ,\eqno(83) \]
then the canonical Hamilton becomes 
\[ \mathcal{H}_c = \frac{1}{h} \Sigma + \left( -\frac{h^1}{h}\right) I\!\!P + \xi^1\phi_1 + \xi\phi + \xi_1\phi^1 \eqno(84) \]
where
\[\hspace{-2.8cm} I\!\!P = p^a f_{,1}^a \eqno(85a) \]
\[ \Sigma = \frac{1}{2} \left[ (p^a)^2 - \Delta(f_{,1}^a)^2\right].\eqno(85b) \]
If eq. (85), $p^a \equiv \frac{\partial\mathcal{L}_f}{\partial f_{,0}^a} = h f_{,0}^a + h^1 f_{,1}^a$ and $\Delta = \det h^{\mu\nu}$.  In addition to the PB algebra of eq. (69), we have 
\[ \left\lbrace \phi_1, \Delta \right\rbrace = \left\lbrace \phi, \Delta \right\rbrace = \left\lbrace \phi^1, \Delta \right\rbrace = 0 \eqno(86a) \]
\[\hspace{-1.5cm} \Delta_{,1} = h\phi^1 + h^{11}\phi_1 - 2h^1\phi \eqno(86b) \]
\[ \left\lbrace \Sigma(x), \Sigma(y) \right\rbrace = (\Delta(x) I\!\!P(x) \partial_1^y - \Delta(y) I\!\!P(y) \partial_1^x) \delta(x-y)\eqno(86c) \]
\[\hspace{.9cm}\left\lbrace \Sigma(x), I\!\!P(y)\right\rbrace = \left( -\Sigma(x) \partial_1^y + \Sigma(y) \partial_1^x + \frac{1}{2} (f_{,1}^a)^2 \Delta_{,1} \right) \delta(x-y) \eqno(86d) \]
and so in addition to the three primary first class constraints $(\Pi_1, \Pi, \Pi^1)$ and the three secondary first class constraints $(\phi_1, \phi, \phi^1)$, there are now just two tertiary first class constraints $(\Sigma, I\!\!P)$.  The generator $G$ of eq. (81) now is extended to become 
\[ G = \int dx^1 \left[ a^1\Pi_1 + a\Pi + a_1\Pi^1 + b^1\phi_1 + b\phi + b_1\phi^1 + c_\Sigma \Sigma + c_{I\!\!P} I\!\!P \right].\eqno(87) \]
Eq. (B.5) now relates $(a^1, a, a_1)$ and $(b^1, b, b_1)$ to $(c_\Sigma, c_{I\!\!P})$ but does not fix them exactly [88]; consequently in eq. (6), the integrals over $c_\Sigma$ and $c_{I\!\!P})$ no longer are integrals over the ``gauge orbit'' associated with all gauge equivalent configurations of the fields. The arbitrariness inherent in the solutions for $(a^1, a, a_1)$ and $(b^1, b, b_1)$ must also be taken into account somehow.  Possibly using in eq. (6) a generator $G$ derived by using the approach of Castellani (eq. (B.10)) would be useful in this circumstance.

An additional problem we would like to address is relating this approach to the path integral in phase space, which is based on the Faddeev-Popov to the path integral in configuration space, to the BFFV approach of refs. [12-18].

\section*{Acknowledgements}
Roger Macleod had numerous helpful suggestions.

\section*{Appendix A. Integration Over Gauge Orbits}

The Faddeev-Popov technique for factoring out the integration over gauge orbits in configuration space [8] can be illustrated by considering the simple example [64, 65, 92]
\[ I = \int d\vec{h} \exp - (\vec{h}^T \mathbf{M} \vec{h} ) \eqno(A.1) \]
where $\vec{h}$ is an $n$-dimensional vector and $\mathbf{M}$ is an $n\times n$ matrix. If the matrix $\mathbf{M}$ has a non-vanishing determinant, then
\[ I = (\pi^n/\det \mathbf{M})^{1/2} . \eqno(A.2) \]
However, if 
\[ \mathbf{M} \mathbf{A}\vec{\theta} = 0 \eqno(A.3) \]
where $\mathbf{A}$ is some matrix and $\vec{\theta}$ is an arbitrary vector, then $\mathbf{M}$ has a vanishing eigenvalue and $\det \mathbf{M} = 0$ so that $I$ is ill-defined. Faddeev and Popov have shown how the ``infinity'' residing in $I$ in this case can be ``factored out'' of $I$ by rescaling $I$ by the divergent integral $\int d\vec{\theta}$. A convenient presentation of the Faddeev-Popov procedure appears in [89].

Initially the form of $I$ in eq. (A.1) is rescaled by the constant factors. 
\[ 1 = \int d\vec{\theta}_1 \delta \left( \mathbf{F} (\vec{h} + \mathbf{M}\vec{\theta}_1)-\vec{p}\right)\det(\mathbf{FM}) \eqno(A.4a)\]
\[ 1 = \int d\vec{\theta}_2 \delta \left( \mathbf{G} (\vec{h} + \mathbf{M}\vec{\theta}_2)-\vec{q}\right)\det(\mathbf{GM}) \eqno(A.4b)\]
and then a further rescaling of $I$ by the constant
\[1 = \pi^{-n} \int d\vec{p}\, d\vec{q}\, e^{-\vec{p}^T\mathbf{N}\vec{q}} \det(\mathbf{N}) \eqno(A.5a) \]
is inserted. If only the constant of eq. (A.4a) were to be used then in place of eq. (A.5a), then we would employ
\[1 = \pi^{-n/2} \int d\vec{p}\, e^{-\vec{p}^T\mathbf{N}\vec{p}} \mathrm{det}^{1/2}(\mathbf{N}). \eqno(A.5b) \]

Eq. (A.1) now becomes, upon integration over $\vec{p}$ and $\vec{q}$, 
\[\hspace{-3cm} I = \pi^{-n} \int d\vec{h} \int d\vec{\theta}_1 d\vec{\theta}_2 \det (\mathbf{FM}) \det (\mathbf{GM}) \det(\mathbf{N}) \eqno(A.6)\]
\[\exp - \left[ \vec{h}^T \mathbf{M} \vec{h} + \left( \mathbf{F} (\vec{h} + \mathbf{M} \vec{\theta}_1)\right)^T \mathbf{N}\left( \mathbf{G}(\vec{h} + \mathbf{M} \vec{\theta}_2)   \right)\right] .\nonumber \]
Upon making the shift $\vec{h} \rightarrow \vec{h} - \mathbf{M} \vec{\theta}_1$ and letting $\vec{\theta} = \vec{\theta}_2 - \vec{\theta}_1$ then eq. (A.6) becomes
\[\hspace{-3cm} I = \pi^{-n} \int d\vec{\theta}_1 \int d\vec{\theta} \int d\vec{h} \det(\mathbf{FM}) \det (\mathbf{GN} ) \det (\mathbf{N} ) \eqno(A.7a) \]
\[ \exp - \left[ \vec{h}^T \mathbf{M} \vec{h} + (\mathbf{F} \vec{h} )^T \mathbf{N} \left(\mathbf{G} (\vec{h} + \mathbf{M} \vec{\theta})\right)\right].\nonumber \]
The integral over $\vec{\theta}_1$ has been ``factored out'' leaving well defined integrals over $\vec{h}$ and $\vec{\theta}$. Exponentiating the determinants $\det(\mathbf{FM})$ and $\det(\mathbf{GM})$ leads to two complex Fermionic ghosts analogous to the usual Faddeev-Popov ghosts while if $\det \mathbf{N}$ is exponentiated a complex Fermionic ``Nielsen-Kallosh'' ghost [6, 90, 91] arises.  The quantity $\vec{\theta}$ in eq. (A.7a) is a ``Bosonic ghost''.

The integral in eq. (A.7a) involves two gauge fixing conditions, $\mathbf{F}\vec{h} = 0$ and $\mathbf{G}\vec{h} = 0$.  Having two conditions is useful when dealing with the transverse-traceless gauge for spin two [64], spontaneously broken gauge theories [65] and gauge theories on a hypersphere [92].  Normally, one only requires a single gauge fixing condition, in which case one would only insert eqs. (A.4a) and (A.5b) into eq. (A.1), leading to the conventional result
\[\hspace{-4cm} I = \pi^{-n/2} \int d\vec{\theta} \int d\vec{h} \det (\mathbf{FM}) \mathrm{det}^{1/2}(\mathbf{N}) \eqno(A.7b) \]
\[ \exp - \left[ \vec{h}^T \mathbf{M} \vec{h} + (\mathbf{F}\vec{h})^T \mathbf{N} (\mathbf{F}\vec{h})\right].\nonumber \]
The integral over the gauge parameter $\vec{\theta}$ again is a divergent multiplicative factor and $\det(\mathbf{FM})$ and $\det^{1/2}(\mathbf{N})$ give rise to a complex Fermionic ghost (the Faddeev-Popov ghost) and a real Fermionic ghost (the Nielsen-Kallosh ghost) respectively.

\section*{Appendix B. The Gauge Generator}

A model with first class constraints has a degree of arbitrariness that is associated with a gauge invariance present in the action [19-27].  In this appendix we consider the generator $G$ of a gauge transformation on the phase space variables $(q_i(t), p^i(t))$ when there are first class constraints $\phi_{a_{i}}$. The index $i$ in $\phi_{a_{i}}$ refers to the ``generation'' of the constraint ($i = 1$, primary; $i = 2$, secondary; 
$i = 3$, tertiary; etc.).  We will first consider gauge invariance in the extended action
\[ S_E = \int_{t_i}^{t_f} dt \left[ p^i \dot{q}_i - H_c(q_i, p^i) - 
\lambda_{a_{i}} \phi_{a_{i}} (q_i, p^i)\right] \eqno(B.1) \]
using the approach of Henneaux, Teitelboim and Zanelli (HTZ) [53, 54]. The generator $G$ of the transformation is taken to be a linear combination of the first class constraints
\[ G =  \mu_{a_{i}}(q_j(t), p^j(t), \lambda_{a_{j}}(t),t) 
\phi_{a_{i}} (q_j(t), p^j(t)) \eqno(B.2) \]
and the change in a dynamical variable $A(q_i(t), p^i(t))$ is taken to be the PB
\[ \delta A = \left\lbrace A, G \right\rbrace .\eqno(B.3) \]
We find that 
\[ \delta S_E = \int_{t_i}^{t_f} dt \left[ \frac{D\mu_{a_{i}}}{Dt} \phi_{a_{i}} + \left\lbrace G, H_c + \lambda_{a_{i}} \phi_{a_{i}} \right\rbrace - \delta \lambda_{a_{i}} \phi_{a_{i}}
\right]. \eqno(B.4) \]
Here $\frac{D}{Dt}$ denotes a time derivative apart from implicit dependence on time through $q_i(t)$ and $p^i(t)$ and $\delta \lambda_{a_{i}}$ is the corresponding change in the Lagrange multiplier 
$\lambda_{a_{i}}$. (Time dependence of $\lambda_{a_{j}}$ through $q_i(t)$ and $p^i(t)$ is cancelled by the PB $\left\lbrace \int_{t_i}^{t_f} dt\, p^i \dot{q}_i, G\right\rbrace$ provided surface terms at $t_i$ and $t_f$ are dropped.)

The extended action $S_E$ reduces to the total action $S_T$ when  $\lambda_{a_{i}} = 0$ for $i \geq 2$.  The total action has the same invariance as the classical action $S_{c\ell} = \int_{t_i}^{t_f} dt L(q_i(t), \dot{q}_i(t))$ [93]. We see from eq. (B.4) that $\delta S_T = 0$ provided 
\[ \frac{D\mu_{a_{i}}}{Dt} \phi_{a_{i}} + \left\lbrace G, H_c + \lambda_{a_{1}} \phi_{a_{1}} \right\rbrace - \delta \lambda_{a_{1}} \phi_{a_{1}} = 0; \eqno(B.5) \]
this entails setting $\lambda_{a_{i}} = \delta\lambda_{a_{i}} = 0$ for $i \geq 2$.  If there are $N$ generations of first class constraints, then eq. (B.5) can be used to fix $\mu_{a_{N-p-1}}$ in terms of $\mu_{a_{N-p}} (p = 0, 1 \ldots N-1)$ by considering the coefficients of $\phi_{a_{i}}\; (i = 2 \ldots N)$ with $\delta\lambda_{a_{1}}$ being fixed by the coefficient of $\phi_{a_{1}}$.  We start by having $\mu_{a_{N}}$ depending solely on $t$. This process need not lead to a unique expression for $G$. A similar approach to finding $G$ appears in ref. [13]. 

In the alternate approach of Castellani (C), the gauge generator is determined by considering the invariances of the Hamilton equations of motion rather than the invariances of the total action [55, 56].  If the equations of motion possess the invariances $q_i \rightarrow q_i + \alpha_i$, $p^i \rightarrow p^i + \beta^i$, then by eq. (B.3)
\[\alpha_i = \left\lbrace q_i, G\right\rbrace = \frac{\partial G}{\partial p^i}\quad \beta^i = \left\lbrace p^i, G\right\rbrace = -\frac{\partial G}{\partial q_i} . \eqno(B.6) \]
The equation of motion $\frac{dA}{dt} \approx \left\lbrace A, H_T \right\rbrace + \frac{\partial A}{\partial t}$ (where $H_T = H_c + \lambda_{a{_1}} \phi_{a{_1}}$) when applied to eq. (B.6) results in 
\[ \dot{\alpha}_i \approx  \left\lbrace \frac{\partial G}{\partial p^i}, H_T \right\rbrace + \frac{\partial^2G}{\partial tp^i}, \quad 
\dot{\beta}^i \approx - \left\lbrace \frac{\partial G}{\partial q_i}, H_T \right\rbrace - \frac{\partial^2G}{\partial tq_i} . \eqno(B.7) \]
The ``weak equality'' $\approx$ in these equations means that they need only be satisfied when the primary constraints vanish.  In addition, the original equations of motion show that 
\[
 \dot{q}_i +\dot{\alpha}_i \approx \frac{\partial}{\partial p^i}
H_T (q_i + \alpha_i, p^i + \beta^i )\nonumber \]
\[ \eqno(B.8) \]
\[\dot{p}^i +\dot{\beta}^i \approx -\frac{\partial}{\partial q_i}
H_T (q_i + \alpha_i, p^i + \beta^i )\nonumber  \]
which becomes to lowest order
\[
\dot{\alpha}_i \approx \frac{\partial}{\partial p^i} \left( \frac{\partial H_T}{\partial q_i} \alpha_i + \frac{\partial H_T}{\partial p^i}\beta^i \right)\nonumber \]
\[ \eqno(B9) \]
\[\dot{\beta}_i \approx -\frac{\partial}{\partial q_i} \left( \frac{\partial H_T}{\partial q_i} \alpha_i + \frac{\partial H_T}{\partial p^i}\beta^i \right). \nonumber \]
Upon equating $\dot{\alpha}_i$ and $\dot{\beta}^i$ in eqs. (B.7) and (B.9) and eliminating $\alpha_i$ and $\beta_i$ by eq. (B.6) we obtain an equation for $G$.  Expanding $G$ in the following way
\[ G = \epsilon(t) G_0 + \dot{\epsilon}(t) G_1 + \ldots + \epsilon^{(N-1)} (t) G_{N-1} \]
if there are $N$ generations of first class constraints, then this equation for $G$ can be satisfied iteratively provided
\[ \epsilon \left\lbrace G_0, H_T \right\rbrace + \dot{\epsilon} \left[ G_0 + \left\lbrace G_1, H_T \right\rbrace \right] + \ddot{\epsilon} \left[ G_1 + \left\lbrace G_0, H_T \right\rbrace  \right] \nonumber \]
\[+\ldots  + \epsilon^{(N-1)} \left[ G_{N-2} + \left\lbrace G_{N-1}, H_T \right\rbrace \right] + \epsilon^{(N)}[ G_{N-1}] \approx 0.\eqno(B.10) \]
Eq. (B.10) is satisfied by taking
\[ G_{N-1} \approx (\mathrm{primary\;constraint})\eqno(B.11) \]
\[ G_{N-2} + \left\lbrace G_{N-1}, H_T \right\rbrace  \approx (\mathrm{primary\;constraint})\nonumber \]
\[ \vdots \nonumber \]
\[  \left\lbrace G_0, H_T \right\rbrace  \approx (\mathrm{primary\;constraint}).\nonumber \]
This permits us to derive the generator of the gauge transformation that leave the action invariant when expressed in terms of phase space variables by examining the Hamilton equations of motion. We are currently considering how to use the generator $G$ found by using eq. (B.11) in conjunction with path integral in phase space.

It should be noted that in using eq. (B.5) to solve for $(\mu_{a_{1}} \ldots \mu_{a_{N-1}})$ in terms of $\mu_{a_{N}}(t)$, a unique solution can only be obtained if the number of first class constraints in each generation is the same.  This is generally the case, though if one considers a scalar field on a two dimensional surface this no longer is true [88].  In this instance there are fewer constraints in the $N^{th}$ generation than in the $(N-1)^{st}$ generation which means that $G$ is not unique.  In general, when using eq. (B.11) to determine $G$, $G$ can be unique only if the number of first class constraints at each generation is the same. 

A unique solution for $G$ may also be contingent on $H_c$ being at most linear in each of the first class constraints. In the four dimensional Palatini action, this is not the case [60].

\section*{Appendix C. Conversion of the Path Integral from Phase Space to Configuration Space}

In this appendix, we will adapt the approach of refs. [57-59] to converting the path integral from phase space to configuration space so that the eq. (6) can be used to quantize models containing a gauge invariance.  We will work with a denumerable number $n$ of degrees of freedom in order to follow refs. [57-59] as closely as possible.

If there are no constraints in the theory, then by eq. (2) [9]
\[ <\mathrm{out}|\mathrm{in}> = \int Dq_i (t) Dp^i (t) e^{i\int dt(p^i\dot{q}_i - H_c(q_i,p^i))}. \eqno(C.1) \]
where $|\mathrm{out}>$ and $|\mathrm{in}>$ are the states associated with the asymptotic values of $q_i(t)$ as $t \rightarrow \pm \infty$.  The equations 
\[p^i = \frac{\partial L(q_i,\dot{q}_i)}{\partial \dot{q}_i} \eqno(C.2)\]
can be solved to yield
\[\dot{q}_i = f_i(q_i,p^i).\eqno(C.3) \]
By using the standard identity for the Dirac delta function 
\[ \delta (f(x)) = \sum_i \delta(x - a_i)/|f^\prime(a_i)| \;\;(f(a_i) = 0) \eqno(C.4) \]
we obtain
\[ \delta (v_i - f_i (q_i, p^i)) = |A_n (q_i,v_i)|\delta \left( p^i - \frac{\partial L(q_i,v_i)}{\partial v_i} \right) \eqno(C.5) \]
where the Hessian matrix is given by 
\[ A_n(q_i,\dot{q}_i) = \frac{\partial^2 L(q_k,\dot{q}_k)}{\partial \dot{q}_i \partial \dot{q}_j} \;\; (i, j = 1 \ldots n). \eqno(C.6) \]
Since upon using eq. (C.3)
\[ H(q_i, p^i) = p^i \dot{q}_i - L(q_i, \dot{q}_i )\eqno(C.7) \]
we can rewrite eq. (C.1) as 
\[ <\mathrm{out}|\mathrm{in}> = \int Dq_i Dp^i Dv_i \delta\left( v_i - f_i(q_i, p^i)\right)  e^{i\int dt(p^i(\dot{q}_i- v_i) + L(q_i,v_i))}. \eqno(C.8) \]
Using eq. (C.5) this becomes
\[ = \int Dq_i Dp^i Dv_i |A_n (q_i, v_i)| \delta \left( p^i - \frac{\partial L (q_i, v_i)}{\partial v_i} \right)  e^{i\int dt(p^i\dot{q}_i - v_i) + L(q_i,v_i))}, \eqno(C.9) \]
or, upon integration over $p_i$ and then shifting $v_i \rightarrow v_i + \dot{q}_i$ we obtain 
\[ <\mathrm{out}|\mathrm{in}> = \int Dq_i \Lambda_n (q_i, \dot{q}_i) 
e^{i\int dt L(q_i,\dot{q}_i)} \eqno(C.10) \]
where
\[\hspace{-7cm}\Lambda_n(q_i, \dot{q}_i) = \int Dv_i |A_n (q_i, v_i + \dot{q}_i)| \eqno(C.11) \]
\[\exp i \int dt \left( L(q_i, v_i + \dot{q}_i) - L(q_i, \dot{q}_i) - v_i \frac{\partial}{\partial v_i} L(q_i, v_i + \dot{q}_i)\right). \nonumber \]
The derivation of eq. (C.10) from eq. (C.1) can be generalized to deal with systems with $N$ generations of first class constraints $\phi_{a_{i}} (q_j, p^j) (i = 1, \ldots, N)$.  Upon choosing a gauge condition $\psi(q_i, \dot{q}_i) = 0$, then by eqs. (2) and (6) we have 
\[ <\mathrm{out}|\mathrm{in}> = \int D\lambda_{a_{i}} Dq_i (t) Dp^i(t) \left[ D\mu_{a_{N}}(t) \delta \left( \psi(q_i, \dot{q}_i) + \left\lbrace \psi (q_i, \dot{q}_i), G(q_i, p^i)\right\rbrace \right.\right. \nonumber \]
\[ \left. \left. -k(t) \right) \Delta (q_i,\dot{q}_i)\right] \exp i \int dt \left[ p^i \dot{q}_i - H_c (q_i, p^i) - \lambda_{a_{1}} \phi_{a_{1}} (q_i, p^i)\right]\eqno(C.12) \]
where $G$ is the generator of gauge transformations
\[ G = \mu_{a_{i}}\phi_{a_{i}} (q_j, p^j) \eqno(C.13) \]
and $\lambda_{a_{i}}$ is a Lagrange multiplier. We are still assuming that the number of first class constraints in each generation is the same and  that $\mu_{a_{1}} \ldots \mu_{a_{N-1}}$ is fixed in terms of $\mu_{a_{N}}$ by eq. (B.5).  If the rank of the Hessian matrix
\[ A_r(q_i, \dot{q}_i) = \frac{\partial^2 L(q_k, \dot{q}_k)}{\partial \dot{q}_i \partial \dot{q}_j} \;\; (i, j = 1 \ldots n) \eqno(C.14) \]
is $r < n$, then there are $n - r$ primary constraints $\phi_{a_{1}}$.  The first $r$ of the equations (C.2) can be solved to give 
\[ \dot{q}_i^\prime = f_i (q^\prime_j, q^{\prime\prime}_j, p^{\prime j}, \dot{q}_j^{\prime\prime}) \;\; (i = 1 \ldots r) \eqno(C.15) \]
and we are left with $n - r$ primary constraints 
\[ \phi_{a_{1}} (q_i, p^i) = p^{i\prime\prime} - g^i (q_j^\prime, q_j^{\prime\prime}, p^{\prime j}) \;\; (i = r + 1 \ldots n) \eqno(C.16) \]
which we take to be first class.  (For canonical invariables $1 \ldots r$ we use a single prime, while for variables $r + 1 \ldots n$ we use a double prime.) 
Employing eqs. (C.7, C.15, C.16) we find that with a gauge function $\psi(q_i, \dot{q}_i)$ and inserting $1$ from eq. (19) 
\[ \hspace{-7cm}<\mathrm{out}|\mathrm{in}> = \int Dq_i\, Dp^i\, Dk \,D\mu_{a_{N}}Dv_{i} \left[ \delta (\psi + \left\lbrace \psi ,G\right\rbrace - k) \Delta \right] \nonumber\]
\[ \hspace{-5.5cm}\delta (v_i - f_i (q_j, p^j, \dot{q}_j^{\prime\prime})) \delta(\phi_{a_{1}} (q_i, p^i))\eqno(C.17) \]
\[\exp i \int dt \left[ p^{i\prime} (\dot{q}_i^\prime - f_i) + (p^{i\prime\prime} - g^i)\dot{q}_i^{\prime\prime} + L(q_i,  v_i, \dot{q}_i^{\prime\prime}) - \frac{1}{2\alpha} k^2 \right]. \nonumber \]
The gauge transformation generated by $-G$ leaves both $\Delta$ and the action invariant, and so eq. (C.17) becomes 
\[ \hspace{-7cm}<\mathrm{out}|\mathrm{in}> = \int D\mu_{a_{N}} \int Dq_i\, Dp^i Dv_i\delta \left(\psi(q_i, v_i, \dot{q}_i^{\prime\prime} \right) - k) \Delta(q_i, v_i, \dot{q}_i^{\prime\prime}) \eqno(C.18) \]
\[ |A_r(q_i, v_i, \dot{q}_i^{\prime\prime})| \delta \left( p^i - \frac{\partial L(q_i, v_i, \dot{q}_i^{\prime\prime})}{\partial v_i}\right)\nonumber \]
\[ \delta (\phi_{a_{1}} (q_i, p^i)) \exp i \int dt \left[  \frac{\partial L(q_i, v_i, \dot{q}_i^{\prime\prime})}{\partial v_i} (\dot{q}_i^\prime - v_i) + L(q_i,  v_i, \dot{q}_i^{\prime\prime}) - \frac{1}{2\alpha} k^2 \right] \nonumber \]
upon using eq. (C.5).  In eq. (C.18), $A_r$ is given by the Hessian matrix
\[ A_r(q_k, \dot{q}_k^\prime, \dot{q}_k^{\prime\prime}) = \frac{\partial^2 L(q_k, \dot{q}_k^\prime, \dot{q}_k^{\prime\prime})}{\partial \dot{q}_i^\prime \partial \dot{q}_j^\prime} \;\; (i,j = 1 \ldots r) \eqno(C.19) \]
which is the analogue of eq. (C.6) for the situation in which there are $n - r$ first class primary constraints.  Upon dropping the (infinite) scale factor coming from integrating over $\mu_{a_{N}}$, integrating over $k$, and making the shift $v_i \rightarrow v_i + \dot{q}_i^\prime$, we are left with eq. (C.19) becoming
\[ <\mathrm{out}|\mathrm{in}> = \int Dq_i\, \Lambda_r (q_i, \dot{q}_i) \exp i \int dt L(q_i, \dot{q}_i) \eqno(C.20) \]
where 
\[ \Lambda_r (q_i, \dot{q}_i) = \int Dv_i\, e^{-\frac{i}{2\alpha} \int dt \left[\psi(q_i, v_i + \dot{q}_i^\prime,\dot{q}_i^{\prime\prime})\right]^2}
\Delta(q_i, v_i + \dot{q}_i^\prime,\dot{q}_i^{\prime\prime})\nonumber \]
\[  |A_r(q_i, v_i + \dot{q}_i^\prime,\dot{q}_i^{\prime\prime})| \delta\left(\phi_{a_{1}} \left(q_i, \frac{\partial L(q_i, v_i + \dot{q}_i^\prime,\dot{q}_i^{\prime\prime})}{\partial v_i},\right.\right.\nonumber \]
\[  \left. \left. g^i\left(q_i, \frac{\partial L(q_i, v_i + \dot{q}_i^\prime,\dot{q}_i^{\prime\prime})}{\partial v_i}\right)\right)\right) \exp i \int dt \bigg[ L(q_i, v_i + \dot{q}_i^\prime,\dot{q}_i^{\prime\prime}) - L(q_i, \dot{q}_i^\prime,\dot{q}_i^{\prime\prime}) \nonumber \]
\[\left. -v_i \frac{\partial L(q_i, v_i + \dot{q}_i^\prime,\dot{q}_i^{\prime\prime})}{\partial v_i} \right].\eqno(C.21)\]

It should be noted that in arriving at this final expression for $<\mathrm{out}|\mathrm{in}>$ we have encountered the PB $\left\lbrace \psi, G\right\rbrace$ appearing in eq. (6).  One might think that this implies having a to compute an ill defined PB such as $\left\lbrace \dot{q}_i(t), p^j(t)\right\rbrace$, but in fact we only encounter this PB in the functional determinant $\Delta$ appearing in eq. (6).  Consequently, time derivatives appearing in the PB can be applied to the Fermionic ghosts used to exponentiate this functional determinant, as in eq. (18).

\end{document}